# Non-Resonant Impulsively Stimulated Raman Scattering by a Terahertz Field: a Case Study of 1$T$-$TaS_2$

Haotian Zhang (张昊天)[1], Yuheng Guo (郭禹恒)[1], Zidu Yu (郁子都)[1], Yongbo Lv (吕涌波)[1], Yiting Wang[2,3], Liwen Feng (冯利文)[1], Jiaying Xu (徐嘉英)[1], Tianlong Xia (夏天龙)[2,3*], Xinbo Wang (汪信波)[4*], Hao Chu (储灏)[1*]

[1]*School of Physics and Astronomy, Shanghai Jiao Tong University, Shanghai 200240, China*

[2]*School of Physics and Key Laboratory of Quantum State Construction and Manipulation (Ministry of Education), Renmin University of China, Beijing 100872, China*

[3]*Laboratory for Neutron Scattering, Renmin University of China, Beijing 100872, China*

[4]*Beijing National Laboratory for Condensed Matter Physics, Institute of Physics, Chinese Academy of Sciences, Beijing 100190, China*

**ABSTRACT**. Time-domain ultrafast and nonlinear terahertz spectroscopy techniques are recently applied to many condensed matter systems for investigating their collective excitations. In centrosymmetric systems, these collective modes are typically Raman-active and therefore do not couple directly to the terahertz electric field. The mechanism by which light-matter interaction realizes in these studies has not been explicitly discussed in detail. In this work, we perform terahertz pump − optical probe and terahertz third harmonic generation investigations on 1$T$-$TaS_2$, a material exhibiting a rich charge-density-wave (CDW) phase diagram including the commensurate, nearly-commensurate and incommensurate CDW phases. The transition between these distinct states leaves a clear signature on the dynamical Raman response. We investigate how the Raman-active phonons couple to a broadband monocycle terahertz field as well as a narrowband multicycle terahertz field. Our results indicate that a modified impulsively stimulated Raman scattering mechanism involving two-photon absorption, also known as non-resonant Raman scattering, underlies the coherent excitation and observation of the lattice modes. These results are relevant for future spectroscopy investigation and coherent control of collective modes using low-energy terahertz field as well as cavity electrodynamical dressing of solids.

## I. INTRODUCTION.

Condensed phases in solids are often characterized by collective excitations that dictate how fast the ordered state fluctuates or responds to external stimuli. In centrosymmetric materials and parity-even ordered phases, these collective modes are typically Raman-active (i.e. they describe even-parity fluctuations of the ordered state and therefore do not possess an electric dipole moment). As a result, these modes do not linearly couple to electromagnetic field in spectroscopy experiments. In recent decades, using ultrafast optical lasers to excite and probe solid state systems, experiments routinely uncover coherent oscillations of various types of collective modes [1–11]. To explain the generation and observation of such coherent dynamics, two mechanisms have been proposed, the impulsively stimulated Raman scattering (ISRS) [12] and the displacive excitation of coherent phonon (DECP) [13] (Fig. 1), specifically under the context of *optical* pump − *optical* probe experiment.

In more recent years, ultrafast terahertz sources become increasingly available and found versatile applications in the study of low-energy dynamics in solids [14–19]. While resonant terahertz excitation demonstrated interesting applications such as nonlinear phononics and possible light-induced superconductivity [20–22], terahertz field off-resonant with IR-active phonons or interband electronic transitions is also shown to have major impacts on solids. Specifically, the dominant interaction mechanism was thought to be the giant Coulomb force resulting from the intense terahertz electric field which drives the free carriers or bound charges to realize a terahertz-driven intraband current or field-induced ionization [23–26]. Examples of such resonant and non-resonant terahertz − matter interaction have been thoroughly reviewed in ref [27].

Interestingly, many recent studies suggest that a moderate low-energy terahertz field (~ several kV/cm) is also able to couple to Raman-active collective modes directly (i.e. not via an intermediate IR-active phonon such as in nonlinear phononics) [14–17]. However, the mechanism by which light − matter interaction takes place on the meV energy scale or the picosecond time scale underlying the terahertz pulse has not been explicitly investigated. To elucidate such mechanism, we perform monocycle terahertz pump − optical reflectivity probe (TPOP) measurement as well as multicycle terahertz drive − third harmonic generation (THG) measurement on the charge-density-wave (CDW) system 1*T*-$TaS_2$. Both measurements suggest that the terahertz field excites the Raman-active mode through two-photon absorption (i.e. instead of their energy difference as in ISRS). This process can be recognized as one type of non-resonant Raman scattering with below-gap optical excitation [28]. Therefore, we refer to this mechanism as non-resonant ISRS. Our elucidation of such mechanism using 1*T*-$TaS_2$ as a material platform has important implications because terahertz excitation and dressing of low-energy collective modes has been recognized as an important approach in engineering novel phases of matter, such has been demonstrated or proposed in many terahertz pump − probe studies [19–22,29] and intrinsic to cavity engineering of low-temperature phases of matter via vacuum fluctuation dressing [30,31]. Therefore, we emphasize that the focus of the present study is not on novel materials physics behind 1*T*-$TaS_2$, but rather the general mechanism of terahertz-matter coupling in centrosymmetric systems.

*Contact author: tlxia@ruc.edu.cn, xinbowang@iphy.ac.cn, haochusjtu@sjtu.edu.cn

## II. RESULTS

We exfoliate high-quality 1*T*-$TaS_2$ single crystals with lateral dimensions reaching one centimeter onto a Kapton tape and a $SiO_2$ substrate, realizing millimeter-sized flakes of thickness varying between ~ 100 nm and a few µm. These two kinds of samples are used for the THG and TPOP measurements respectively. The monocycle terahertz pulse, generated from $LiNbO_3$ using the tilted pulse front technique, realizes a peak electric field of 1 MV/cm and a half-cycle of ~ 0.5 ps. To produce the narrowband multicycle terahertz pulse, the monocycle pulse is passed through two bandpass filters of 0.42 THz (0.7 THz) central frequency. This allows subsequent THG measurements at 1.26 THz (2.1 THz) frequency, which is facilitated by two additional bandpass filters at the THG frequency after the sample. These filters improve the dynamical range of THG measurement by suppressing the 0.42 THz (0.7 THz) linear driving field which otherwise dominates the terahertz transmission. The transmitted terahertz electric field profile is measured by electro-optical sampling in a 1 mm ZnTe crystal. Ultrafast optical reflectivity change is measured with 35 fs pulses of 800 nm central wavelength using two balanced Si photodiodes.

1*T*-$TaS_2$ transitions from a nearly-commensurate CDW (nc-CDW) state to a commensurate CDW (c-CDW) state as it cools down through 170 K [32,33]. In the c-CDW state, an energy gap (> 200 meV) in the electronic structure opens up at the Fermi level [33,34], resulting in a reduced DC electric conductivity. In terms of AC conductivity or terahertz reflectivity, a similar decrease is expected, implying that terahertz transmission increases in the c-CDW state. This is indeed manifested in the temperature dependence of the 0.42 and 0.7 THz linear transmission through the sample (Fig. 2).

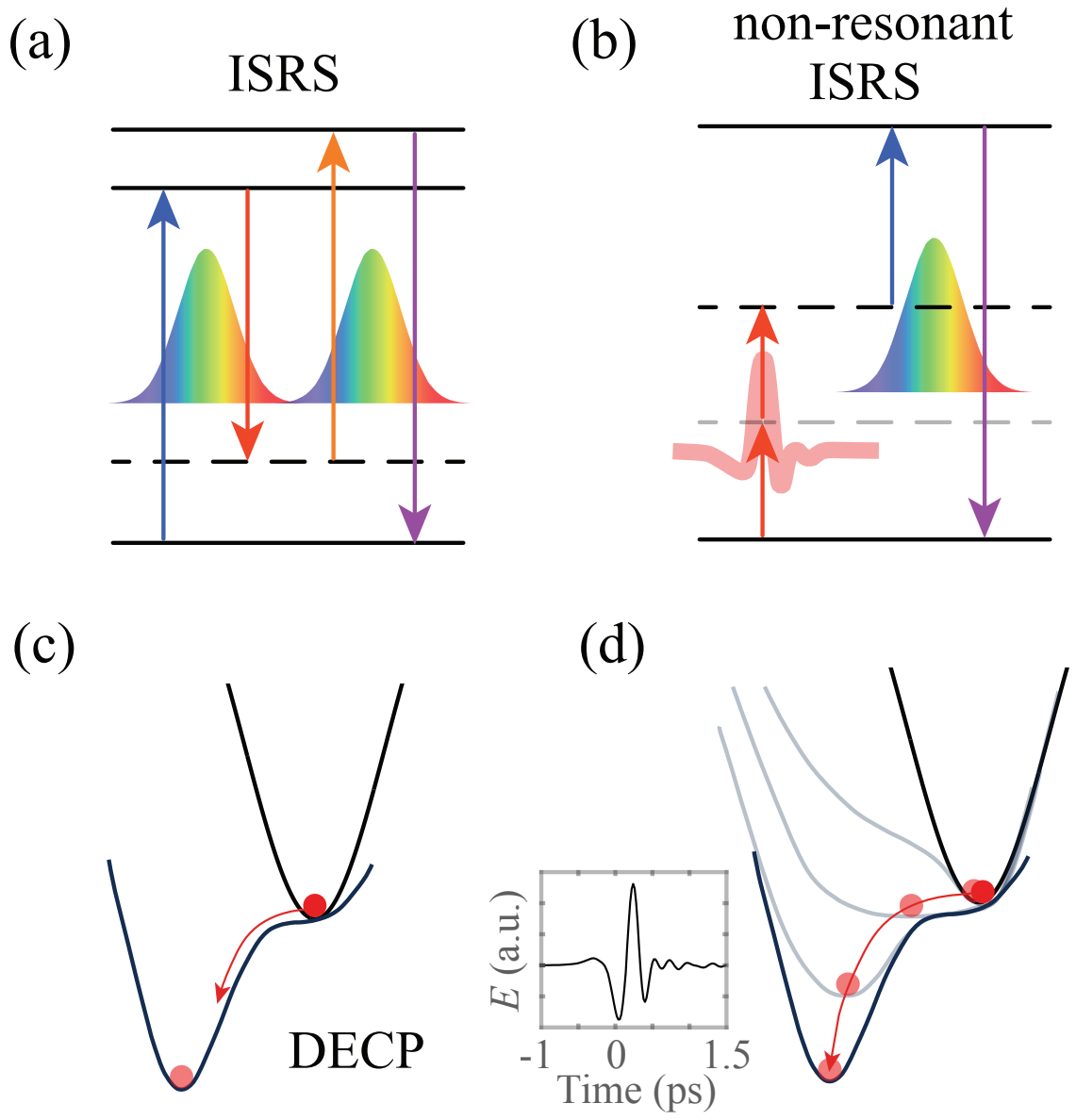

**Figure 1 Mechanisms for generating coherent Raman-active modes in solids (a)** In Impulsively Stimulated Raman Scattering (ISRS), two photons from the excitation pulse creates a stimulated Raman excitation via their energy difference. A time-delayed probe pulse undergoes inelastic scattering with the Raman excitation, resulting in an anti-Stokes shifted Raman photon that interferes with the elastically scattered photons, which manifests as a beating, or intensity oscillations of the reflected probe beam. **(b)** The proposed non-resonant ISRS mechanism for generating coherent Raman-active modes using low-frequency terahertz excitation. Here, two terahertz photons coherently add to create the Raman excitation. A third photon, of either optical frequency or the same terahertz frequency, scatters inelastically with the Raman excitation, resulting in coherent oscillations in optical reflectivity and terahertz third harmonic generation. **(c)** In Displacive Excitation of Coherent Phonons (DECP), an ultrafast optical excitation results in a spatial redistribution of electron charge. Therefore, the electrostatic environment in which the atoms reside is suddenly modified. The atoms respond to the sudden change by coherently moving to the new minimum in the electrostatic potential, resulting in coherent phonon oscillations. **(d)** If the electrostatic potential is modified on a timescale similar to the phonon's timescale, the lattice can respond fast enough to the modification and follow suit. Therefore, no coherent oscillation of the lattice on similar timescale as the perturbation is expected. The waveform of the monocycle terahertz field used for the experiment is shown in the inset.

A notable hysteresis between heating and cooling is also observed in this temperature window due to the first-order nature of the nc-CDW-to-c-CDW transition, consistent with DC conductivity results. A unique triclinic CDW (t-CDW) state appears above the c-CDW state and below the nc-CDW state during heating and not during cooling (Fig. 2). As both the t-CDW and nc-CDW states retain a finite Fermi surface, the transition between them does not leave a clear signature on DC conductivity. In our measurement, terahertz linear transmission shows no discontinuity between these two states either.

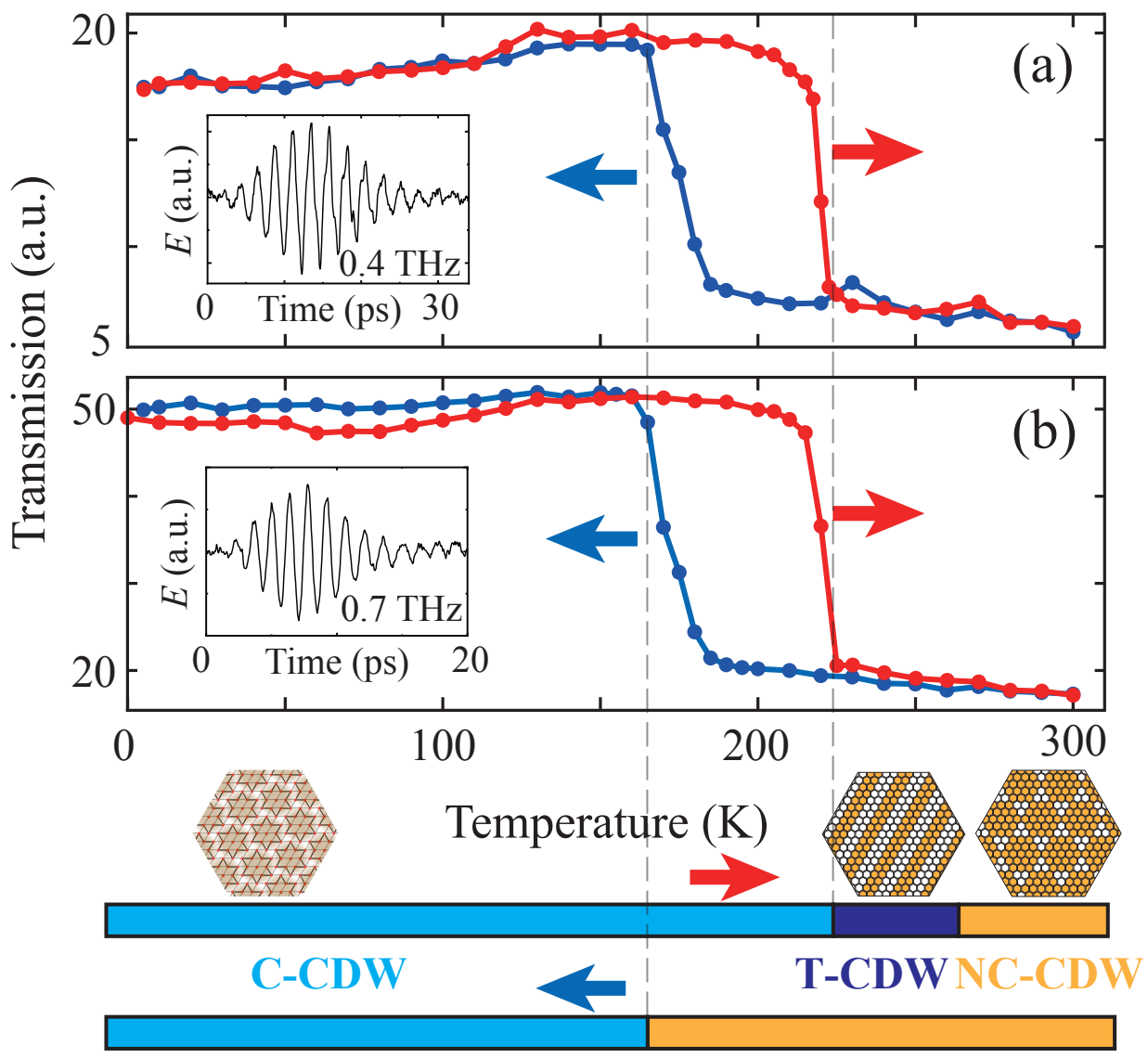

**Figure 2 Terahertz linear transmission through 1*T*-$TaS_2$ (a)** 0.42 THz and **(b)** 0.7 THz linear transmission through 1*T*-$TaS_2$ as a function of temperature. The cooling and heating cycle shows a hysteresis between 160 K and 220 K, due to the hysteretic charge-density-wave (CDW) transitions in this material as illustrated below. Insets show the multicycle 0.42 THz and 0.7 THz waveforms used for these measurements.

Having established the expected c-CDW transition in our sample, next we perform monocycle terahertz pump − 800 nm reflectivity probe measurement on the sample on $SiO_2$ substrate. The inset of Fig. 3(a) shows a representative ultrafast reflectivity transient measured at 10 K. Multiple sets of long-lived coherent oscillations are visible in the raw data in the c-CDW state, resulting from the folding of $\boldsymbol{q}_{CDW}$ phonons to the Brillouin zone center by the long range CDW ordering [35]. We perform fast Fourier transform on these oscillations (after subtracting a smooth background) and plot the temperature-dependent FFT spectra as a 2D colormap in Fig. 3(b). Above-noise level spectral peaks are observed at 2.05, 2.37, 2.65, 2.95, 3.12 THz frequencies (see Supplementary Information for results from another sample showing additional peaks at 2.15, 3.38, 3.81 THz). Some of these oscillations (2.05 THz, 3.12 THz, all $A_g$ modes) have been reported in previous 800 nm pump − 800 nm probe studies [5,11], while others (2.37 THz, 2.65 THz, 2.95 THz) are uniquely seen in our experiment (see Supplementary Information for discussion). On approaching the c-CDW transition temperature ($T_{CDW}$) ~ 220 K, the 2.38 THz phonon softens towards the 2.05 THz phonon, indicating that this mode might be strongly coupled to the c-CDW order parameter.

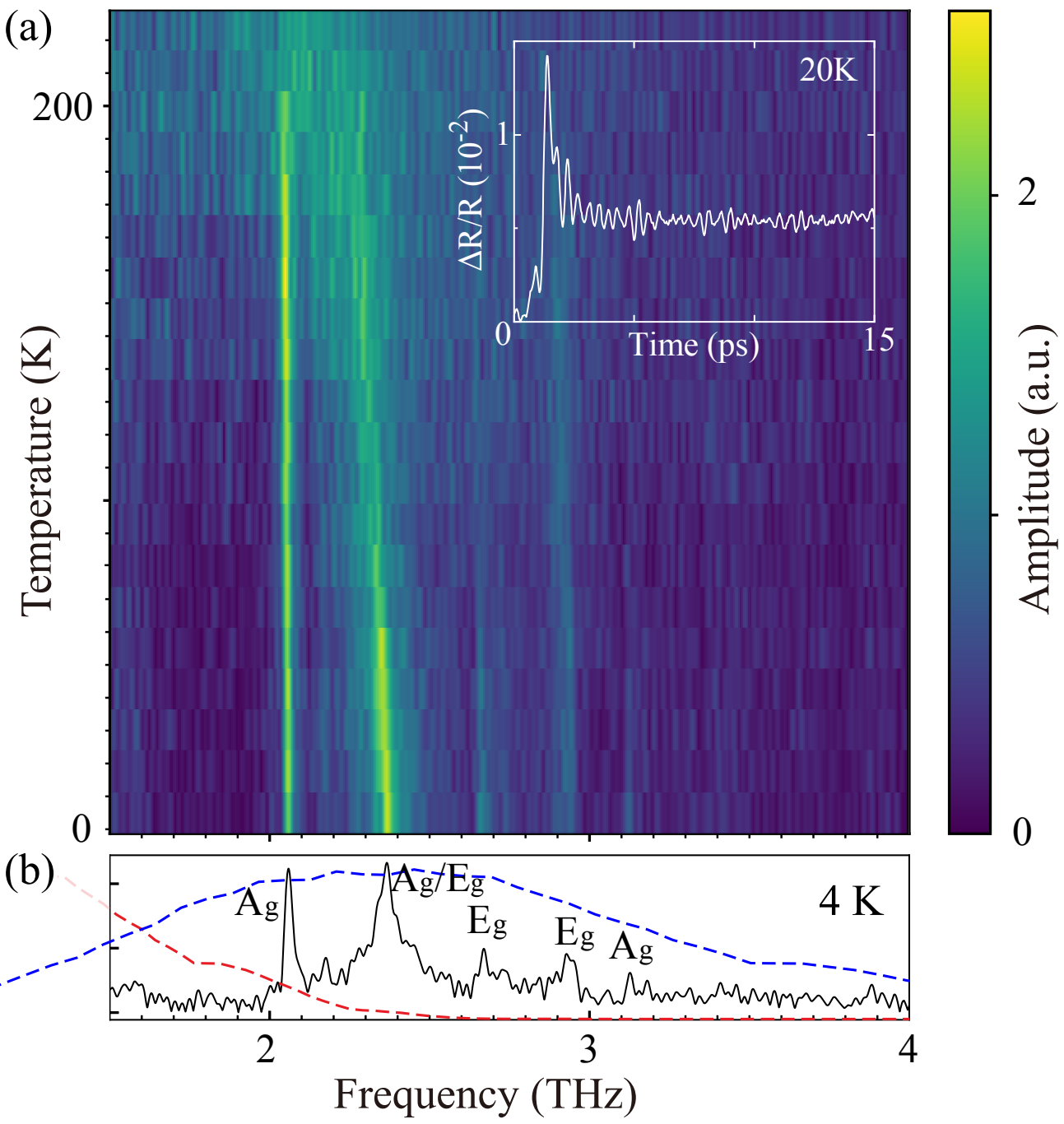

**Figure 3 Coherent Raman-active phonons launched by a broadband terahertz excitation (a)** The broadband monocycle terahertz pulse launches a series of Raman-active phonons in 1*T*-$TaS_2$ in the c-CDW state, with frequencies as high as 3.1 THz (also see Supplementary Information for a different sample with a 3.4 THz mode). **(b)** The phonon spectrum at 4 K with each mode's symmetry labeled according to polarization-resolved equilibrium Raman scattering measurement. Also shown is the power spectrum of the broadband terahertz pulse used for this measurement (red dashed line), and the same spectrum scaled by a factor of ×2 in frequency (blue dashed line). The observed Raman-active modes all fall within twice the terahertz excitation bandwidth.

While all these modes find matches in equilibrium Raman scattering measurement, here the surprising fact is that the 0.5 ps-long terahertz field is able to launch long-lived coherent oscillations up to 3.12 THz (3. 81 THz for the sample in Supplementary Information), which falls outside the bandwidth of the excitation pulse. We note that a recent TPOP measurement on another van der Waals material $FePS_3$ reports a remarkably large-amplitude and long-lived 7.5 THz phonon oscillation launched by a broadband terahertz pulse with little spectral weight at the phonon frequency [19]. In both our and their studies, there is no evidence that these Raman-active phonons are coupled to the terahertz field via an intermediate IR-active phonon. To explain these surprising results, we propose that these Raman excitations are created by two-photon absorption of the terahertz pulse, as illustrated in Fig. 1(b). This mechanism enables Raman-active modes with frequency as high as twice the terahertz bandwidth to be excited, which is consistent with our results and the findings in ref [19]. This also represents a modification of the ISRS mechanism, where two photons from the same excitation pulse undergoes difference frequency generation to create the Raman excitation.

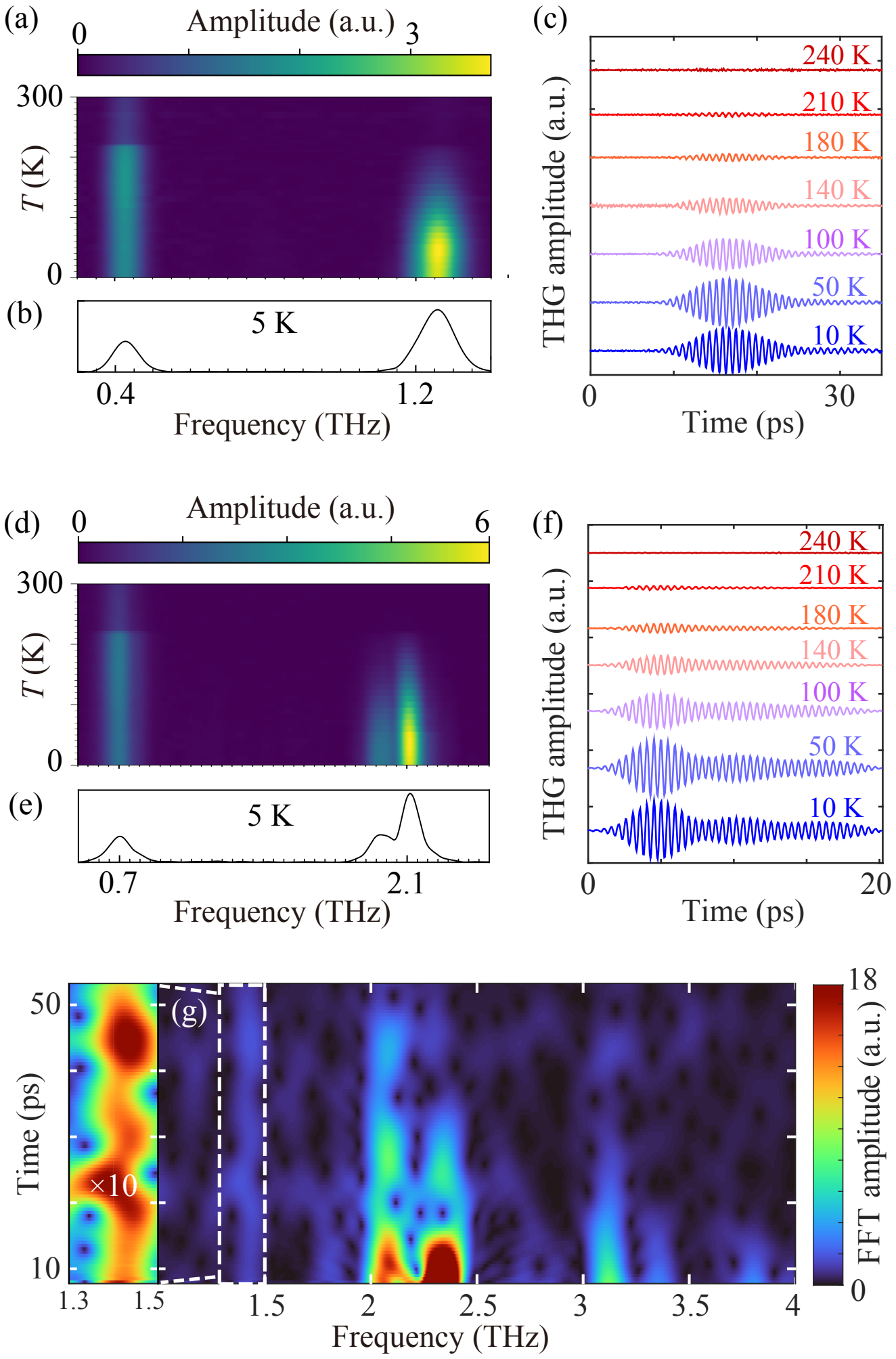

**Figure 4 Terahertz third harmonic generation in 1*T*-$TaS_2$** **(a)** Temperature dependence of THG amplitude under 0.42 THz drive and **(b)** representative Fourier amplitude spectrum at 5 K. **(c)** Extracted-THG waveforms at representative temperatures across the c-CDW transition. **(d)-(f)** Corresponding results obtained with 0.7 THz drive. A notable side peak develops next to the main THG peak in the Fourier spectrum. The existence of the side peak is corroborated by the beating-like pattern in the THG waveforms. **(g)** Coherent phonon spectrum obtained with 15 ps moving-time window FFT analysis on optical pump − optical reflectivity transients measured at 80 K. A periodic modulation of the coherent phonon amplitude is observed for the 2.05 THz and 2.37 THz mode as well as a 1.4 THz mode (intensified and zoomed-in on the left).

Here, we additionally argue that DECP is unlikely to be the mechanism behind the coherent phonon generation in our experiment and in ref [19]. As illustrated in Fig. 1(c), DECP requires an ultrafast modification of the electrostatic potential surrounding the atoms on a timescale much faster than the lattice dynamics, so that the atoms respond to the sudden change by co-ordinated movement towards the new potential minimum. This is typically achieved through resonant interband excitation using optical or mid-IR pump, as dictated by the energy-scale of the interband electric dipole moment, which is often faster than the envelope of the excitation pulse. In comparison, our broadband terahertz spectrum falls far below the CDW energy gap, excluding any possibility of resonant excitation even by multiphoton absorption. In addition, the monocycle terahertz pulse has an electric field cycle of ~ 0.5 ps, which is comparable to the timescale of the observed coherent phonons. The terahertz field-induced modification of the electrostatic potential, as felt by the atoms in their respective normal coordinates, is therefore gradual and similar to the situation illustrated in Fig. 1(d). Given the slow evolution of the electrostatic potential, it is unlikely that long-lived coherent oscillations such as the 3.81 THz phonon (see Supplementary Information) can be launched by the DECP mechanism. Lastly, we also note that the additional modes observed in our study are all $E_g$ phonons as previously identified in polarization-resolved Raman scattering measurements [36]. According to the original formulation of DECP [13], only $A_g$ phonons can be excited, providing an additional argument that the non-resonant ISRS mechanism underlies the excitation of these modes.

To provide a further perspective, we also apply narrowband terahertz excitation to the sample exfoliated on Kapton tape and demonstrate how the multicycle terahertz field interacts with the collective modes. A significant terahertz THG response is observed in the c-CDW state with a hysteretic temperature dependence similar to terahertz linear transmission. Below the c-CDW transition, the THG intensity displays a temperature dependence consistent with an increasing CDW order parameter (Fig. 4(a)(d)), indicating that the THG process couples directly or indirectly to the CDW order parameter. In the 1st directly coupled scenario, the CDW amplitude mode (usually underdamped) or phase mode (usually heavily damped) may participate in the two-photon scattering process. In the 2nd indirectly coupled scenario, CDW-coupled phonons may participate in the two-photon scattering process. There exists a 3rd scenario in which a terahertz field-depinned sliding CDW [37–39] (sometimes distinguished from a pure CDW amplitude mode or phase mode [38]) participates in the 3rd one-photon interaction (i.e. direct absorption) process illustrated in Fig. 1(b), regardless of the nature of the two-photon scattering process beforehand.

Our data show that while the 1.26 THz THG spectrum manifests a symmetric Gaussian profile, the 2.1 THz THG spectrum exhibits a side peak around 1.97 THz (Fig. 4(b)(e)). These spectral features are respectively corroborated by the Gaussian-enveloped THG waveform in Fig. 4(c) and a beating-like THG response in Fig. 4(f). To explain the appearance of the side peak under 0.7 THz drive, we note that optical pump − optical reflectivity probe measurements in ref [11] (see their Supporting Information) reveals a periodic modulation of the amplitude of a 1.43 THz phonon as a function of pump – probe delay. This possibly indicates a periodic/back-and-forth energy transfer (i.e. coherent coupling) between this mode and other degree(s) of freedom in the system. In our own 1200 nm pump – 800 nm reflectivity probe measurements (Fig. 4(g)), we similarly observe periodic amplitude modulation of this mode as well as the 2.05 and 2.37 THz mode as a function of pump – probe delay. Corroborating this observation with the THG results above, we speculate that the side peak in the 2.1 THz THG spectrum arises because the 0.7 THz field quadrtically couples to the 1.43 THz Raman-active phonon, whose coherent coupling yet to other degree(s) of freedom in the system leads to periodic modulation of its own amplitude and therefore the THG magnitude. In contrast, the 0.42 THz field probably couples to a continuum of low-energy interlayer Raman-active phonons in the c-CDW state falling between 10 and 60 $cm^{-1}$ (0.3 and 1.8 THz) as predicted by density functional theory study [35], which mostly undergo incoherent scattering (which explains why they don't show up in optical pump − optical reflectivity probe measurements). Therefore, we do not observe a prominent side peak in the THG spectrum or a beating-like THG temporal response. Finally, we also perform driving field dependence measurements to test the 3rd mechanism for THG discussed above. Our measurements down to 20 kV/cm (see Supplementary Information), permitted by the experimental set-up and the beamtime, do not yet show indications for a threshold electric field. We argue that this does not necessarily rule out the 3rd scenario. Instead, it indicates that the threshold field for dislocating a disorder-pinned CDW in our sample is lower than 20 kV/cm.

Finally, we note that two-photon absorption has been a familiar concept to the field of nonlinear optics and solid-state spectroscopy. However, its intuitive picture has been associated with optical photons with quantized energy and particle-like description, whereas low-energy terahertz field is often recognized for its classical wave-like nature. Terahertz two-photon excitation of a Raman-active phonon was reported in diamond nearly a decade ago [16]. The narrowband terahertz field employed in that work centers around 20 THz (~ 15 μm) in the mid-IR range and the Raman-active phonon is at 40 THz, both approaching optical frequencies. In the present study, we employ broadband and multicycle terahertz fields both approaching sub-THz frequency range and show that a modified ISRS mechanism is still valid despite the classical picture of the electromagnetic wave at such frequencies. Importantly, the frequency range investigated in our work is relevant for the collective modes of many interesting condensed phases including superconductivity, CDW, nematicity, etc. By elucidating how terahertz field, either broadband or narrowband, couples to such low-energy collective modes, we envisage that future studies, including both terahertz pump studies and cavity engineering leveraging electromagnetic vacuum fluctuations, will take advantage of the non-resonant ISRS mechanism for coherent control or dressing of such collective modes. In addition, this mechanism may also be exploited for investigating quasi-elastic Raman modes or soft modes associated with symmetry-breaking transitions, which is currently beyond the capability of many conventional Raman scattering setups.

*Acknowledgments* Hao Chu acknowledges support by the National Key Research and Development Program of China (Grants No. 2024YFA1408701), the National Natural Science Foundation of China (Grants No. 12274286) and the Yangyang Development Fund. Xinbo Wang acknowledges support by the National Key Research and Development Program of China (Grants No. 2024YFA1611300) and the National Natural Science Foundation of China (Grants No. 12574349). Tianlong Xia acknowledges support by the National Key Research and Development Program of China (Grants No. 2024YFA1409002). This work was supported by the Synergetic Extreme Condition User Facility (SECUF, https://cstr.cn/31123.02.SECUF)

---

# Supplementary Information for

## Non-Resonant Impulsively Stimulated Raman Scattering by a Terahertz Field: a Case Study of 1*T*-$TaS_2$

### S1. Observation of $E_g$ modes in Our Measurements

The c-CDW state of 1$T$-$TaS_2$ has $C_{3i}$ point group symmetry, for which the Raman tensors for $A_g$ and $E_g$ modes are [1]

$$R_{A_g} = \begin{pmatrix} a & 0 & 0 \\ 0 & a & 0 \\ 0 & 0 & b \end{pmatrix},$$

$$R_{{}^1E_g} = \begin{pmatrix} c & d & e \\ d & -c & f \\ e & f & 0 \end{pmatrix},$$

$$R_{{}^2E_g} = \begin{pmatrix} d & -c & -f \\ -c & -d & e \\ -f & e & 0 \end{pmatrix}.$$

In our monocycle terahertz pump – 800 nm reflectivity probe measurement, the terahertz field is linearly polarized. In this geometry, a subset of $A_g$ and $E_g$ modes ({a, c, d}) is expected to be excited. In fact, in many previous optical pump – probe reflectivity experiments on 1$T$-$TaS_2$, the observed coherent phonons all have $A_g$ symmetry. Our study reveals additional $E_g$ modes previously missing from optical pump – probe studies. We argue that these could be the additional $E_g$ modes ({e, f}) that require an additional out-of-plane electric field to excite. The monocycle terahertz pump pulse used in our study has a full-width-half-maximum diameter of ~ 50 mm. It is focused onto the sample using a 50 mm-diameter parabolic mirror with a focal distance of 50 mm, realizing a numerical aperture (NA) of ~ 0.45. The large NA effectively realizes an out-of-plane electric field component to the terahertz pulse at the focal point, which excites the additional $E_g$ modes observed in our study [2].

We assign the 2.37 THz (79.1 $cm^{-1}$), 2.65 THz (88.4 $cm^{-1}$), 2.95 THz (98.4 $cm^{-1}$) modes to $E_g$ symmetry according to the Raman polarization dependence of ref [3]. In particular, in Fig. 2 of ref [3], around 80 $cm^{-1}$ there are two modes overlapping with each other which peak respectively near 82 $cm^{-1}$ in the parallel-polarized channel ($A_g$ symmetry) and near 80 $cm^{-1}$ in the cross-polarized channel ($E_g$ symmetry). We assign our 2.37 THz mode to the 80 $cm^{-1}$ mode with $E_g$ symmetry.

## S2. Coherent Phonon Spectrum from Another 1*T*-$TaS_2$ Sample

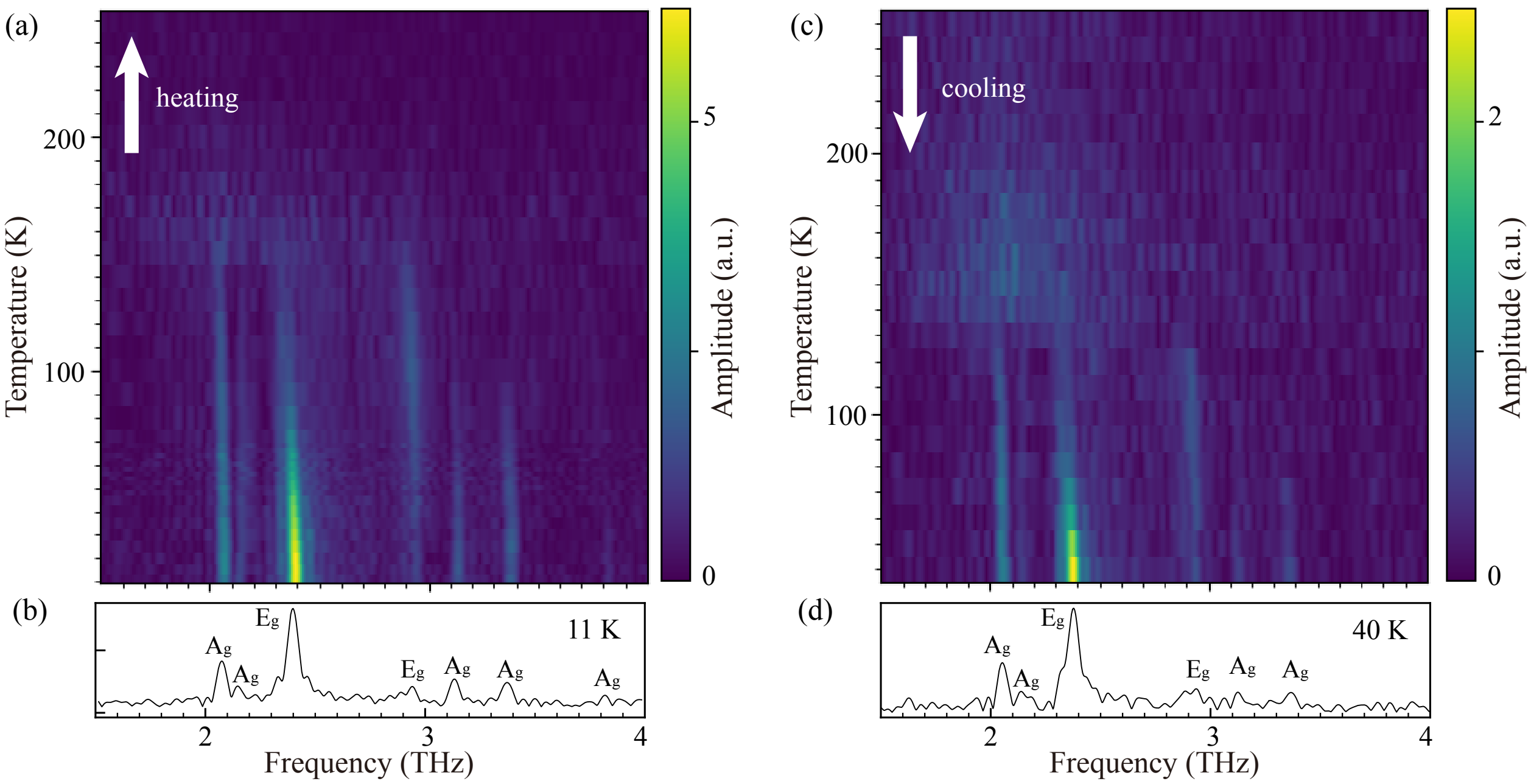

**Figure S1 Temperature dependence of coherent phonon spectrum from TPOP measurements on another 1*T*-$TaS_2$ sample exfoliated on $SiO_2$ substrate. (a)(b)** During the heating process, additional phonon modes at 2.15 THz ($A_g$ mode), 3.38 THz ($A_g$ mode) and 3.81 THz ($A_g$ mode) are observed. **(c)(d)** The sample is then measured while cooling down to 40 K.

We measured another 1*T*-$TaS_2$ sample exfoliated on $SiO_2$ substrate as shown in Fig. S2. Based on the phonon spectrum, the c-CDW transition temperature appears to be ~ 50 K lower in this sample compared to the sample investigated in the main text. This could be due to the different sample thickness, which is known to affect the CDW transition temperature [4], or due to the poor thermal contact between the sample and the substrate in this particular device.

## S3. Driving Field Dependence of Coherent Phonons

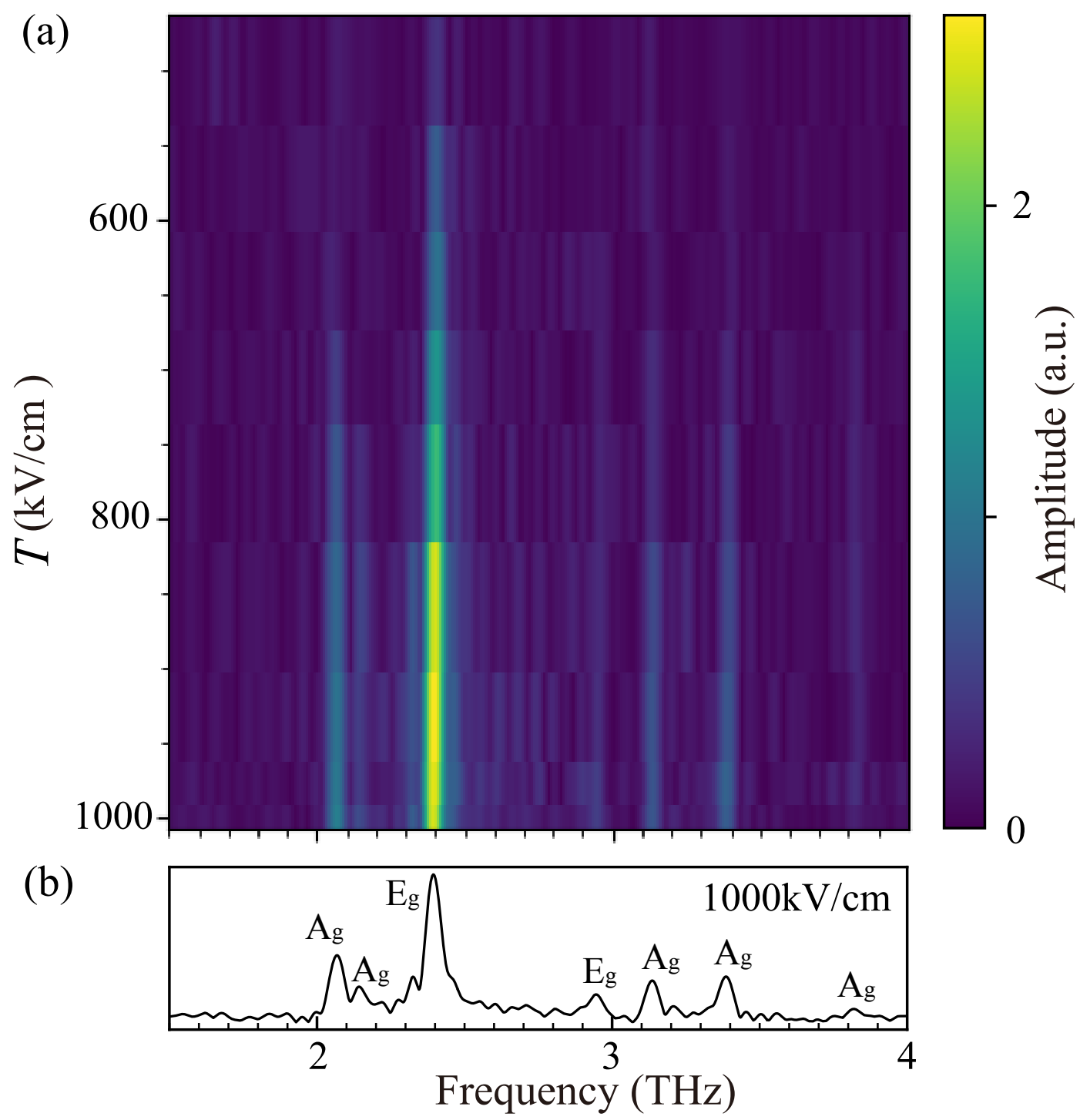

**Figure S2 Coherent phonon spectrum as a function of monocycle terahertz field strength.**

On the second sample studied in Fig. S1, we vary the monocycle terahertz peak electric field between 1 MV/cm and 500 kV/cm and perform optical reflectivity measurement at 11 K. The resulting coherent phonon spectra are shown above.

## S4. Driving Field Dependence of Terahertz Third Harmonic Generation

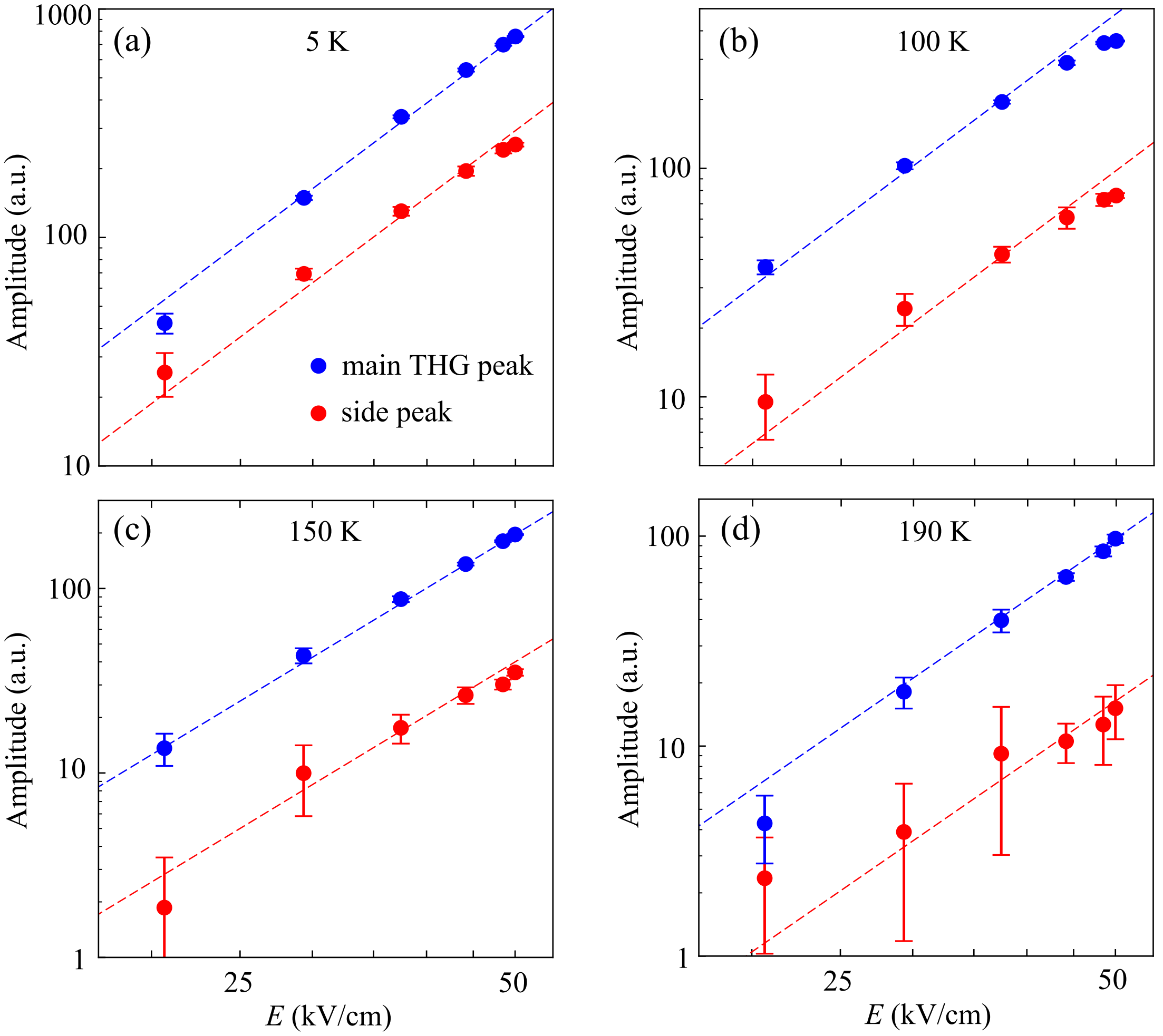

**Figure S3 Driving field dependence of THG response of the c-CDW state of 1*T*-$TaS_2$.** The 0.7 THz incident field is varied between 20 kV/cm and 50 kV/cm peak field strength and the THG response of the c-CDW state is measured at **(a)** 0 K **(b)** 100 K **(c)** 150 K **(d)** 190 K. Dotted lines are cubic power laws as a guide for the eye.

We perform incident field dependence measurement of the THG response under 0.7 THz periodic drive. As temperature approaches the c-CDW transition ~ 220 K during heating, the pinning of the CDW by lattice defects is expected to become weaker because the CDW correlation length decreases. If the THG response arises from a terahertz field-depinned sliding CDW, we expect to see a decrease of the depinning field (i.e. threshold field for generating THG) as temperature increases [5]. However, in the four sets of measurements presented here, we do not see clear evidence of such critical field down to the lowest measurable THG response.

## S5. Phonon Spectrum from Optical Pump – Optical Probe Measurement

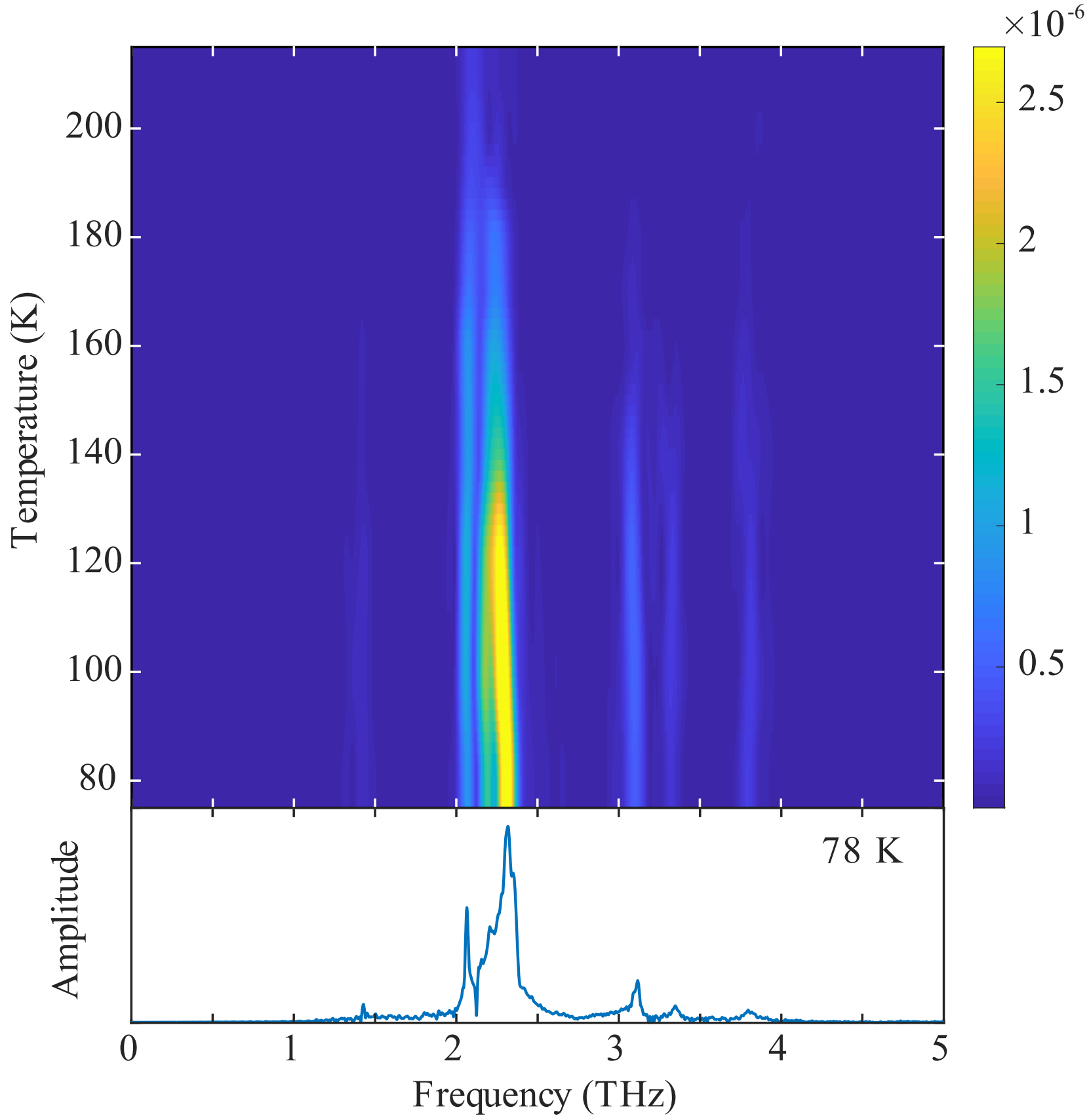

**Figure S4 Coherent phonon spectrum from optical pump – optical reflectivity probe measurements.** Coherent phonons around 1.4, 2.05, 2.15, 2.37, 3.12, 3.38, 3.8 THz are observed.

We perform 1.2 μm pump – 800 nm reflectivity probe measurements on 1*T*-$TaS_2$ thin film. The lower panel in Fig. S4 shows a high-quality coherent phonon spectrum obtained from averaging many scans with long pump – probe time delays at 78 K. A 1.4 THz phonon is clearly identified in the spectrum. The upper panel in Fig. S4 shows the temperature dependence of these phonons, where the 1.4 THz mode persists to ~ 150 K before being significantly broadened and vanishing into background, consistent with ref [6].

## S6. 10 ps Moving-Time Window FFT Analysis of Optical Pump – Optical Probe Results

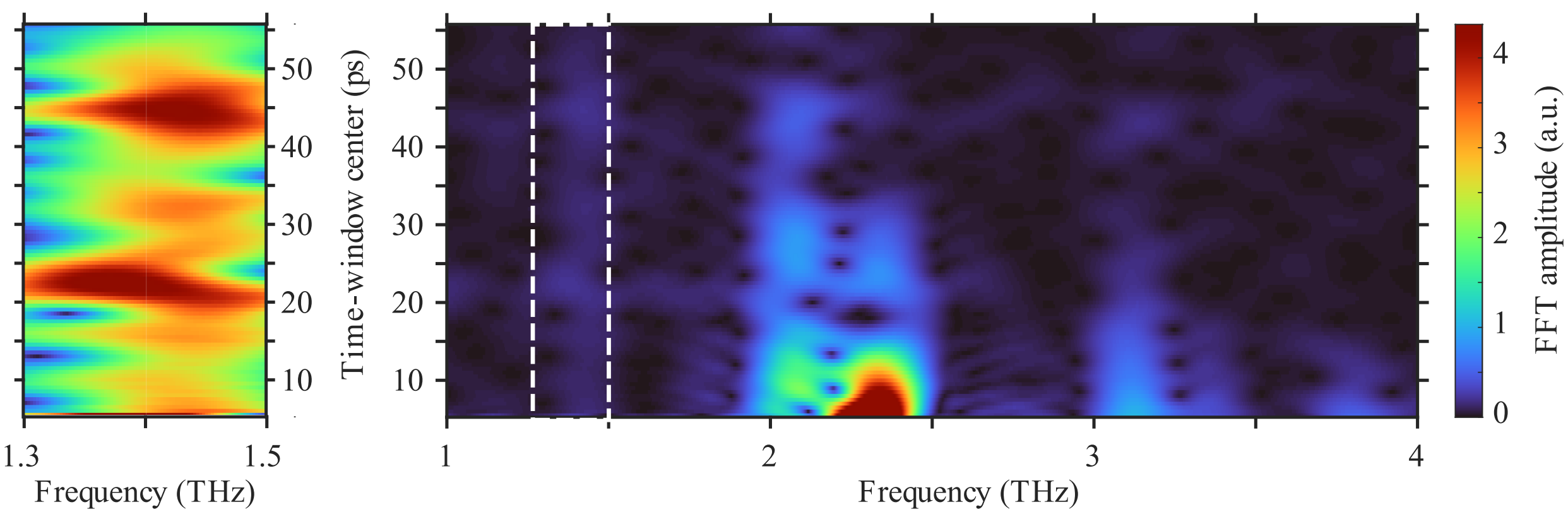

**Figure S5 10 ps moving-time window FFT analysis of coherent phonons from optical pump – optical reflectivity probe measurements.** The frequency range between 1.3 THz and 1.5 THz is zoomed in and plotted with intensified color scale on the left.

To show consistency between moving-time window FFT analysis of coherent phonons using different time window sizes, we analyze the same dataset behind Fig. 4(g) of the main text using a 10 ps moving-time window. The results are shown in Fig. S5. Similar modulations of the 1.4, 2.05 and 2.37 THz phonon amplitude as a function of pump – probe delay time is observed, with similar modulation period as in the main text.